\documentclass[twocolumn]{aastex631}

\usepackage{color}
\usepackage{subfigure}
\usepackage{hyperref}

\newcommand{\cms}{\mbox{cm s$^{-1}$}}
\newcommand{\kms}{\mbox{km s$^{-1}$}}

\newcommand{\thetaLD}{$2.019  \pm  0.012 $} 
\newcommand{\VisLD}{$0.983  \pm  0.011$} 
\newcommand{\alphaLD}{$0.14  \pm  0.03$} 
\newcommand{\thetaUD}{$1.979  \pm  0.006 $} 
\newcommand{\VisUD}{$0.981  \pm  0.011$} 
\newcommand{\radiusLD}{$0.793  \pm  0.004 $} 

\newcommand{\massdv}{$0.780  \pm  0.012$} 
\newcommand{\massvmax}{$0.800  \pm 0.008 $} 
\newcommand{\masslogg}{$0.69  \pm  0.09 $} 

\newcommand{\vsini}{$0.1  \pm  0.1 $} 
\newcommand{\incMWO}{$5  \pm  5 $}

\newcommand{\teff}{$5320 \pm 40$ }
\newcommand{\logg}{$4.48 \pm 0.05$} 
\newcommand{\parallax}{$273.8097 \pm 0.1701$}
\newcommand{\distance}{$ 3.652 \pm 0.003$}  
\newcommand{\lumin}{$0.45 \pm 0.02$}
\newcommand{\prot}{$46 \pm 4$}

\accepted{July 18, 2023}

\submitjournal{AJ}

\shorttitle{$\tau$~Ceti Stellar Parameters}
\shortauthors{Korolik et al.}

\begin{document}

\title{Refining the Stellar Parameters of $\tau$~Ceti: a Pole-on Solar Analog}

\newcommand{\ycaa}{Yale Center for Astronomy and Astrophysics, Yale University, 46 Hillhouse Avenue, New Haven, CT 06511, USA}
\newcommand{\yale}{Department of Astronomy, Yale University, 52 Hillhouse Avenue, New Haven, CT 06511, USA}
\newcommand{\sfsu}{Department of Physics and Astronomy, San Francisco State University, 1600 Holloway Avenue, San Francisco, CA 94132, USA}
\newcommand{\lowell}{Lowell Observatory, 1400 W. Mars Hill Road, Flagstaff, AZ 86001, USA}
\newcommand{\yalephysics}{Department of Physics, Yale University, 217 Prospect St, New Haven, CT 06511, USA}
\newcommand{\ucr}{Department of Earth and Planetary Sciences, University of California, Riverside, CA 92521, USA}
\newcommand{\mira}{Monterey Institute for Research in Astronomy, 200 8th St, Marina, CA 93933, USA}
\newcommand{\flatiron}{Center for Computational Astrophysics, Flatiron Institute, Simons Foundation, 162 Fifth Avenue, New York, NY 10010, USA}
\newcommand{\tsu}{Center of Excellence in Information Systems, Tennessee State University, Nashville, TN 37209, USA}
\newcommand{\michigan}{Department of Astronomy, University of Michigan, Ann Arbor, MI 48109, USA}
\newcommand{\michiganclasp}{Department of Climate and Space Sciences and Engineering, University of Michigan, Ann Arbor, MI 48109, USA}
\newcommand{\exeter}{Astrophysics Group, Department of Physics \& Astronomy, University of Exeter, Stocker Road, Exeter EX4 4QL, UK}
\newcommand{\grenoble}{Institut de Planetologie et d’Astrophysique de Grenoble UGA/CNRS, Grenoble F-38058, France}
\newcommand{\chara}{The CHARA Array of Georgia State University, Mount Wilson Observatory, Mount Wilson, CA 91203, USA}

\correspondingauthor{R.\ M.\ Roettenbacher}
\email{rmroett@umich.edu}

\author{Maria Korolik}
\affiliation{\yale}

\author[0000-0002-9288-3482]{Rachael M.\ Roettenbacher}
\affiliation{\michigan}
\affiliation{\yale}

\author[0000-0003-2221-0861]{Debra A.\ Fischer}
\affiliation{\yale}

\author[0000-0002-7084-0529]{Stephen R.\ Kane}
\affiliation{\ucr}

\author[0000-0002-6703-5406]{Jean M.\ Perkins}
\affiliation{\mira}

\author[0000-0002-3380-3307]{John D.\ Monnier}
\affiliation{\michigan}

\author[0000-0001-9764-2357]{Claire L.\ Davies}
\affiliation{\exeter}

\author[0000-0001-6017-8773]{Stefan Kraus}
\affiliation{\exeter}

\author[0000-0002-0493-4674]{Jean-Baptiste Le Bouquin}
\affiliation{\grenoble}

\author[0000-0002-2208-6541]{Narsireddy Anugu}
\affiliation{\chara}

\author[0000-0002-3003-3183]{Tyler Gardner}
\affiliation{\michigan}
\affiliation{\exeter}

\author[0000-0001-9745-5834]{Cyprien Lanthermann}
\affiliation{\chara}

\author[0000-0001-5415-9189]{Gail H.\ Schaefer}
\affiliation{\chara}

\author[0000-0001-5980-0246]{Benjamin Setterholm}
\affiliation{\michigan}
\affiliation{\michiganclasp}

\author[0000-0002-9873-1471]{John M.\ Brewer}
\affiliation{\sfsu}

\author[0000-0003-4450-0368]{Joe Llama}
\affiliation{\lowell}

\author[0000-0002-3852-3590]{Lily L.\ Zhao}
\affiliation{\flatiron}

\author[0000-0002-4974-687X]{Andrew E. Szymkowiak}
\affiliation{\yale}

\author[0000-0003-4155-8513]{Gregory W. Henry} 
\affiliation{\tsu}

\begin{abstract}

To accurately characterize the planets a star may be hosting, stellar parameters must first be well-determined.   $\tau$~Ceti is a nearby solar analog and often a target for exoplanet searches.  Uncertainties in the observed rotational velocities have made constraining $\tau$~Ceti's inclination difficult.  For planet candidates from radial velocity (RV) observations, this leads to substantial uncertainties in the planetary masses, as only the minimum mass ($m \sin i$) can be constrained with RV.
In this paper, we used new long-baseline optical interferometric data from the CHARA Array with the MIRC-X beam combiner and extreme precision spectroscopic data from the Lowell Discovery Telescope with EXPRES to improve constraints on the stellar parameters of $\tau$~Ceti.  Additional archival data were obtained from a Tennessee State University Automatic Photometric Telescope and the Mount Wilson Observatory HK project.
These new and archival data sets led to improved stellar parameter determinations, including a limb-darkened angular diameter of \thetaLD~mas and rotation period of \prot~days.  By combining parameters from our data sets, we obtained an estimate for the stellar inclination of $7\pm7^\circ$.  This nearly-pole-on orientation has implications for the previously-reported exoplanets.  An analysis of the system dynamics suggests that the planetary architecture described by \citet{feng2017} may not retain long-term stability for low orbital inclinations. 
Additionally, the inclination of $\tau$~Ceti reveals a misalignment between the inclinations of the stellar rotation axis and the previously-measured debris disk rotation axis ($i_\mathrm{disk} = 35 \pm 10^\circ$).

\end{abstract}

\keywords{G dwarf stars (556), long baseline interferometry (932), spectroscopy (1558), stellar properties (1624), solar analogs (1941)}

\

\section{Introduction} 
\label{intro}
Due to its similarity and proximity to the Sun, $\tau$~Ceti (HD 10700) has been studied extensively since the early 1900's \citep[e.g., the parallax observations of][]{adams1916}. Moreover, the star has been of particular interest because it is thought to host planets near its habitable zone \citep{feng2017}. Understanding planet-hosting stars well is critical, as improved stellar parameters can lead to more accurate planetary parameters. \par

$\tau$~Ceti is an inactive, $4.4 - 12.4$ Gyr \citep{lachaume1999, pijpers2003, difolco2004, mamajek2008, baum2022}, G8V \citep{keenan1989} star $3.652 \pm 0.002$ pc away from Earth \citep{gaia2022}. It was selected as one of the first radial velocity (RV) standard stars \citep{tuomi2013}. \citet{feng2017} suggested that $\tau$~Ceti hosts four or more planets detected through RV, two of which are reported to be located near the star's the habitable zone, as defined by \citet{kopparapu2014}. These planets range in mass (as $m \sin i$, where $m$ is the planet's actual mass and $i$ is the \emph{orbital} inclination) between $1.75 - 3.93 \ \mathrm{M}_\oplus$, in orbital period between $20 - 636$ days, and in separation from the star between $0.133 - 1.33$ AU. \par

$\tau$~Ceti has a debris disk that spans approximately  10 to 50 AU \citep{macgregor2016}, with a dust mass of around $1.2 \ \mathrm{M}_\oplus$ \citep{greaves2004}. Planetary formation models imply that the disk and the star share a common plane, with aligned rotation axes.
In previous studies, the inclination of $\tau$~Ceti itself was determined to be $0-40^\circ$ \citep{greaves2004} using the projected rotational velocity from \citet{saar1997} with the stellar rotation period and radius \citep{saar1997, difolco2004}. A  high-angular-resolution study with the Herschel Space Observatory revealed the debris disk of $\tau$~Ceti has an inclination of $35 \pm 10^\circ$ \citep{lawler2014}, in contrast to nearly edge-on results in previous studies with lower-resolution observations  \citep[e.g.,][]{watson2011,greaves2004}. 
\par
Also a target of asteroseismic studies, the detected pulsations of $\tau$~Ceti and similar stars are stochastically excited due to internal convection zones \citep{handler2013}. The pulsation modes are excited over a range of frequencies generally following a normal distribution. They are often described by $\nu_\mathrm{max}$, the frequency of maximum power, and $\Delta\nu$, the frequency difference between consecutive modes of the same angular degree. The $\Delta\nu$ and the $\nu_\mathrm{max}$ are used to determine stellar characteristics such as mass, radius, and evolutionary state. Asteroseismic and stellar parameters are related through scaling relations that allow for unknown parameters to be reliably determined \citep{kjeldsen1995}. $\tau$~Ceti has previously been found to have a $\nu_\mathrm{max} = 4100 \ \mu$Hz and a $\Delta\nu = 169 \ \mu$Hz \citep{teixeira2009}. \par
In this paper, we calculated characteristic stellar parameters of $\tau$~Ceti. We used data on $\tau$~Ceti from interferometry to determine its angular diameter and from spectroscopy to constrain effective temperature, surface gravity, and rotational velocity.  We then combined those values to calculate $\tau$~Ceti's mass. Using an age estimate, we determined a rotation period and compared it to rotation periods derived with new and archival data.   From the rotation period, we determined the stellar inclination and investigated its implications on the orbital stability of $\tau$~Ceti's potential planets.

\section{Observations} 

\subsection{MIRC-X Interferometry}
\label{obs:interferometry}
Long-baseline optical interferometric data were gathered over eight nights, UT 2021 November 2 through November 9, at the Center for High Angular Resolution Astronomy (CHARA) Array.  All six telescopes of the CHARA Array with baselines spanning 34-330m \citep{brummelaar2005} were used on 2021 November 3-7. On November 2, the E1 telescope was not used, and on November 8 and 9, the S2 telescope was not used. The light was combined with the Michigan InfraRed Combiner-eXeter (MIRC-X) beam combiner.   MIRC-X operates in the $H$-band ($\sim 1.6~ \mu$m) and was used with a grism \citep[$R \sim 190$;][]{anugu2020}, as $\tau$~Ceti is very bright \citep[$H = 1.72$;][]{ducati2002}.  We used the standard MIRC-X reduction pipeline (version 1.3.3) and default parameters with the exception of the following parameters (values used are noted in parentheses):  number of coherent co-adds (10),  flux threshold  (5),  signal-to-noise threshold  (3), maximum integration time in seconds for a single data file (150).   The longest and shortest wavelength channels were removed from the data because they were often outliers. The data were median-filtered over five neighboring spectral channels, reducing the number of data points but improving the data quality.   The data were then calibrated with a version of the previous MIRC software \citep{monnier2012} modified to work with MIRC-X data. The calibration stars\footnote{ The star HD 1921 was additionally observed on 2021 November 5-7 as a calibration star.  These data were not used to calibrate $\tau$~Ceti, as it is not a good calibration star because many of its closure phases vary from $-20^\circ$ to $+20^\circ$ (by contrast, the other calibrators have closure phases that mostly vary between $-5^\circ$ to $+5^\circ$).}  used can be found in Table \ref{calibrators}.    \par 

\begin{deluxetable*}{l c c c c}
\label{calibrators}
\tablecaption{MIRC-X Observing Details}
\tablehead{
 \colhead{UT Date} & \colhead{Observing sequence} & \colhead{Angular diameter (mas)} & \colhead{Limb-darkening coefficient ($\alpha$)} & \colhead{Visibility at origin ($V_0$)}
 }
\startdata
 \hline
 2021 Nov 2 & HD 9562 - $\tau$~Ceti - HD 16569 & 2.009 & 0.10 & 0.998 \\
 2021 Nov 3$^*$ & HD 9562 - $\tau$~Ceti - HD 16569 & 2.078 & 0.28 & 0.936\\
 2021 Nov 4 & HD 9562 - $\tau$~Ceti - HD 16569 & 2.034 & 0.19 & 1.017 \\
 2021 Nov 5 & $\tau$~Ceti - HD 16569 & 2.038 & 0.13 & 0.984 \\
 2021 Nov 6 & $\tau$~Ceti - HD 16569 & 2.050 & 0.21 & 1.009 \\
 2021 Nov 7 & $\tau$~Ceti - HD 9562 & 1.997 & 0.13 & 0.966 \\
 2021 Nov 8 & HD 9562 - $\tau$~Ceti - HD 16569 & 2.045 & 0.21 & 1.003 \\
 2021 Nov 9 & HD 9562 - $\tau$~Ceti & 2.046 & 0.17 & 0.986 \\
 All nights$^\dag$ & -- & 2.019 $\pm$ 0.012 & 0.14 $\pm$ 0.03 & 0.983 $\pm$ 0.011 \\
 \hline
\enddata
\tablenotetext{}{The angular diameter for HD 9652 is $\theta_\mathrm{LD} = 0.588 \pm 0.014$~mas and HD 16659 is $\theta_\mathrm{LD} = 0.645 \pm 0.042$~mas from \cite{chelli2016}. \\
$^*$Calibrating this night with only HD 16569 yields $\theta_\mathrm{LD} = 2.065$ mas, $\alpha = 0.23$, and $V_0 = 0.925$.  The all-nights fit using one or both calibrators from Nov 3 are nearly identical with differences in less than one-fifth of the 1-$\sigma$ errors.  \\ 
$^\dag$Best-fit values using all nights and standard deviations from 1000 bootstraps of the entire data set. }
\end{deluxetable*}

\subsection{EXPRES Spectroscopy}
\label{obs:expres}
The Extreme PREcision Spectrograph (EXPRES) at the 4.3m Lowell Discovery Telescope (LDT) run by Lowell Observatory was used to obtain 200 spectra of $\tau$~Ceti over the period of time from 2019 August  to 2021 October. The data from EXPRES reach a median resolving power of $R \sim 137, 500$ and an RV precision of $30$ \cms\ for main-sequence FGK stars with a target signal-to-noise ratio of $\sim250$. The standard EXPRES pipeline was used for reductions \citep{blackman2020, petersburg2020}. The full RV data set is included in Table \ref{expresdata}.  Following \citet{brewer2016}, the standard EXPRES pipeline and the Spectroscopy Made Easy (SME) method \citep{piskunov1996} provided stellar parameters including the effective temperature, $T_\mathrm{eff}$; surface gravity, $\log g$; and rotational velocity, $v \sin i$, as well as each spectrum's RV.  
The errors stated for the stellar parameters are only based upon the variations observed in the spectra during these nights.  This method and its limitations are discussed in \citet{brewer2016}.

\begin{deluxetable}{c c c}
\label{expresdata}
\tablecaption{EXPRES RV Data}
\tablehead{
 \colhead{MJD} & \colhead{RV (m/s)} & \colhead{RV Error (m/s)}
 }

\startdata
\hline
58710.460 & -0.215 & 0.429 \\
58710.461 & -1.067 & 0.373 \\
58710.463 & -1.731  & 0.401 \\
58711.491 & 1.431 & 0.511 \\
58711.492 & 0.264 & 0.520 \\
58711.493 & 0.0267  & 0.468 \\
58712.490 & -0.239 & 0.386 \\
58712.491 & -0.656 & 0.394 \\
58712.493 & -0.108 & 0.404 \\
58714.495 & -0.325 & 0.355 \\
\enddata
\tablenotetext{}{The table is available in its entirety in machine-readable form.}
\end{deluxetable}

\subsection{Mount Wilson Observatory HK Project}
From 1967 through 1995, the Mount Wilson Observatory (MWO) HK Project obtained 1784 S-index measurements for $\tau$~Ceti.  The S-index is a measure of photon counts for the Ca II H and K (in emission for active stars) compared to two nearby continuum bands \citep[for further information, see][]{vaughan1978}.  This value will trace the motion and/or evolution of active regions on the stellar surface.  Details on the data acquisition and analysis can be found in \citet{wilson1968,wilson1978,vaughan1978,duncan1991,baliunas1995}.
 We made use of the 1995 NSO version of the data.  

\subsection{Automatic Photoelectric Telescope Photometry}
Ground-based photometric data were obtained with the T4 0.75m Automatic Photoelectric Telescope (APT) at Fairborn Observatory, AZ from 1996 November 4 through 2020 January 23 \citep{henry1999}. Differential magnitudes were obtained through Str\"omgren \textit{b} and \textit{y} filters and combined into a single $(\textit{b}+\textit{y})/2$ passband. The comparison stars used were HD 10453 and HD 9061, which show no evidence of variation on short or long timescales. Long-term signals were removed from the $\tau$~Ceti data set prior to our analysis.  The trend was determined by applying a Gaussian smoothing to the light curve with a window of 100 days, a value chosen to preserve trends within a rotation period, but remove those across an observing season.  These data were previously published in  \citep{zhao2022}.

\section{Stellar Parameter Determination} 

We find the stellar parameters listed in Table \ref{parameters} with the data described above and from literature values.  The methods and results are described in this section.

\begin{deluxetable*}{l c  c}
\label{parameters}
\tablecaption{Stellar Parameters}
\tablehead{
 \colhead{Parameter} & \colhead{Value} & \colhead{Source}
 }
\startdata
 \hline
Uniform disk diameter, $\mathrm{\theta_\mathrm{UD}}$ (mas) & \thetaUD  & This work\\
Visibility at origin (UD), $V_0$ & \VisUD & This work\\
Limb-darkened disk diameter, $\mathrm{\theta_\mathrm{LD}}$  (mas)  & \thetaLD & This work\\
Visibility at origin (LD), $V_0$ & \VisLD & This work\\
Limb-darkening coefficient, $\alpha$ & \alphaLD & This work\\
Parallax, $\pi$ (mas) & \parallax   & \citet{gaia2022} \\
Distance, $d$ (pc) & \distance & \citet{gaia2022} \\
Radius, $R$ ($R_\odot$) & \radiusLD & This work\\
Rotational velocity, $v \sin i$ (\kms) & \vsini &  This work\\
Effective temperature, $T_\mathrm{eff}$ (K) & \teff & This work\\
Surface gravity, $\log g$ & \logg & This work\\
Mass, $M$ ($M_\odot$) & \masslogg & This work\\
Large frequency separation, $\Delta \nu$ ($\mu$Hz) & 169 & \citet{teixeira2009} \\
Frequency of maximum power, $\nu_\mathrm{max}$ ($\mu$Hz) & 4100 & \citet{teixeira2009} \\
Age, $t$ (Gyr) & 10 & \citet{difolco2004} \\
Rotation period, $P_\mathrm{rot}$ (days) & \prot &This work \\
Inclination, $i$ ($^\circ$) & $7 \pm 7$ & This work\\
 \hline
\enddata
\tablecomments{All parameters based on the angular diameter use the limb-darkened disk diameter. 
}
\end{deluxetable*}

\subsection{Angular Diameter}
\label{analysis:angular diameter}
$\tau$~Ceti is resolved with the CHARA Array. To determine the angular diameter for the model that best matched the interferometric data, the lowest reduced $\chi^2$ between the observations and a model with varying angular diameter was identified. We measured the star's angular diameter by finding the best fit of the model to the observed visibilities.  Modeling the star as a uniform disk, the squared normalized visibility amplitude, $V^2$, is: 
\begin{equation}
\label{eqn:model}
    V^2(B_{\bot}, \lambda, \theta) = \left(\frac{2J_1(\pi\theta B_{\bot}/\lambda)}{\pi\theta B_{\bot}/\lambda}\right)^2,
\end{equation}
 where $J_1$ is the Bessel function of the first order of the first kind with the argument including the angular diameter, $\theta$, the projected baseline, $B_\bot$, and the wavelength of observation, $\lambda$. \par
   \par

\begin{figure}
    \includegraphics[scale=0.55]{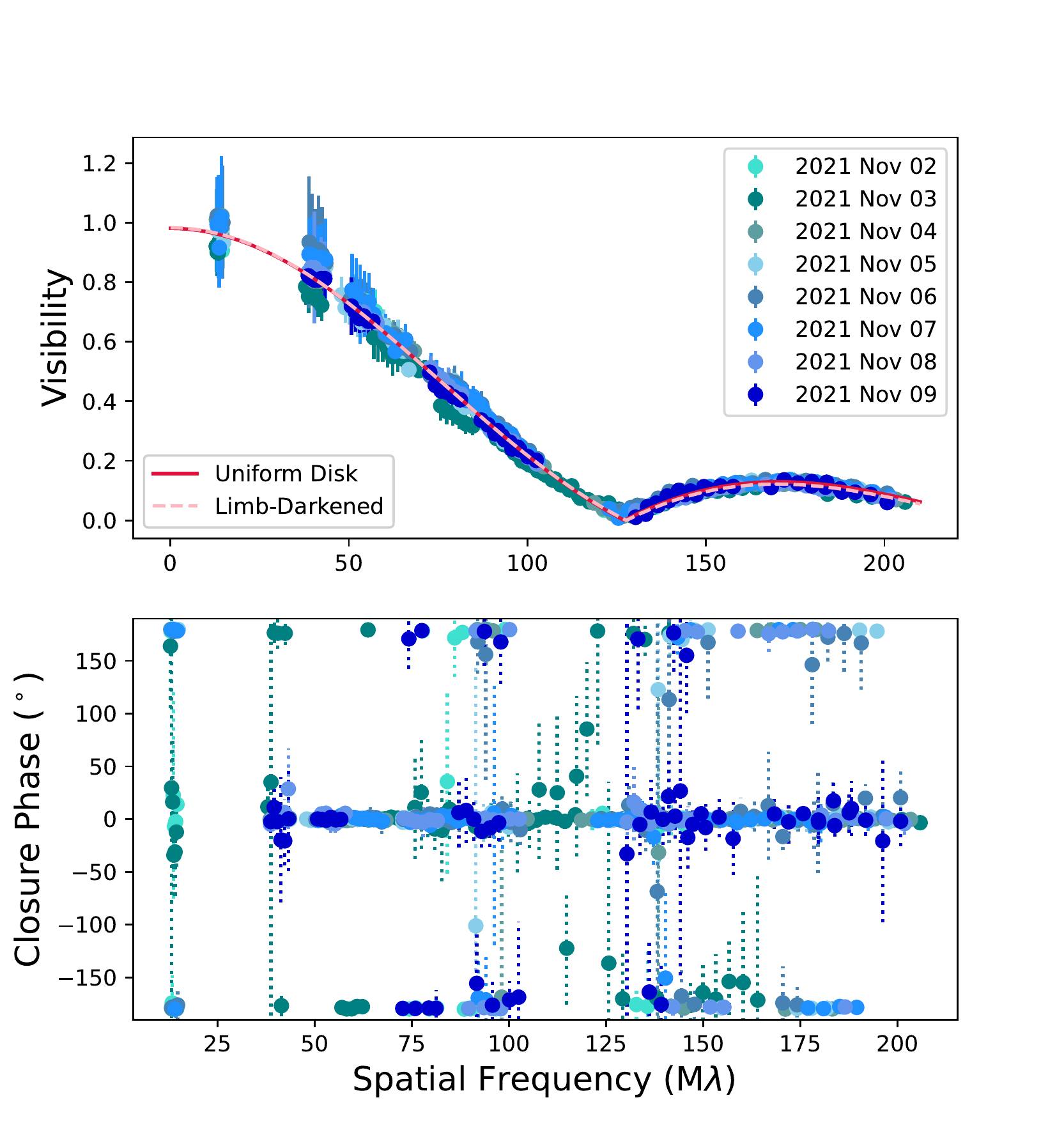}
    \caption{Plot (top) of normalized visibility amplitude versus spatial frequency ($B_{\bot}/\lambda$). The data combines all eight nights of observation for a total of 613 data points. Plot (bottom) of closure phases for all eight nights of observation, showing non-zero phases.
    } 
    \label{visiblity amplitude}
\end{figure}

The data and best fit to the model (Equation \ref{eqn:model}) are included in Figure \ref{visiblity amplitude} and Table \ref{calibrators}. From this analysis, the uniform disk angular diameter of $\tau$~Ceti was determined to be $\theta_\mathrm{UD} =$\thetaUD~mas and the visibility amplitude at a spatial frequency of 0 was $V_0 =$ \VisUD. The errors were determined using a bootstrap for all eight nights of data combined.  That is, the total number of points were chosen from the observations randomly, with replacement 1,000 times.  The errors reported are the standard deviations from those 1,000 iterations.

As a star is not expected to be a uniform disk, but should exhibit limb-darkening, the data were also fit to a power-law limb-darkened model,

\begin{equation}
    I(\mu) = I_0\mu^\alpha,
\end{equation}
where $I$ is intensity, $I_0$ is the intensity at the center of the stellar disk, $\mu$ is the cosine of the angle from the observer to the normal to the stellar surface, and $\alpha$ is the limb-darkening coefficient.  \citet{hestroffer1997} showed that this modifies the visibility amplitude, $V$, to be

\begin{equation}
    \label{eqn:LD}
    V = (\alpha + 2) \int_0^1 \! (1-r^2)^{\alpha/2} \ J_0 (rB_{\bot}/\lambda) \ r \, \mathrm{d}r,
\end{equation}
where $J_0$ is the Bessel function of the zeroth order of the first kind and $r$ is the fractional radius of the star. \par

Fitting for the angular diameter, we determine it to be $\theta_\mathrm{LD}$ = \thetaLD~mas with $V_0 =$ \VisLD~mas and $\alpha = $ \alphaLD.  The value for $\alpha$ is consistent with values reported by \citet{kervella2017} for similar stars: $\alpha$ Centauri A (G2V), $\alpha = 0.1404 \pm 0.0050$; $\alpha$ Centauri B (K1V), $\alpha = 0.1545 \pm 0.0044$; and the Sun, $\alpha = 0.15027$.  

For both the uniform and limb-darkened disks, the values presented here have had a factor of $1.0054 \pm 0.0006$ divided from them, in accordance with a scaling found by \citet{gardner2022} and an update by J.\ Monnier (private communication).   

Both our uniform disk result (\thetaUD~mas) and our limb-darkened result (\thetaLD~mas) are within the range of previous literature values, seen in Table \ref{literature}. 
Discrepancies are likely due to the amount or quality of the data used in the analyses.  The measurement given here used significantly more data than those from the literature, both due to using all six CHARA Array telescopes and multiple nights of observation. \par

Using the Gaia parallax of $\pi =$ \parallax~mas \citep[distance, $d=$ \distance~parsecs (pc);][]{gaia2022}, we determine $\tau$~Ceti has a radius of $R= $~\radiusLD~$R_\odot$.  \par
In the following calculations, we use the limb-darkened disk angular diameter and resultant radius estimate.   \par

\subsection{Temperature}

We analyzed all of the EXPRES spectra following the procedure of \citet{brewer2016} to derive abundances and global stellar parameters, including  $T_{\mathrm{eff}}$, $\log{g}$, metallicity, rotational broadening, and projected rotational velocity ($v \sin{i}$), along with abundances for a few $\alpha$-elements.  In this first stage, other abundances are scaled solar values.  We then perturb the resulting temperature by $\pm 100$~K and re-fit.  The global parameters from the weighted mean of the three models are fixed while abundances for 15 elements are fit.  This new abundance pattern is adopted and the above two steps are repeated to get a final model.

From the EXPRES spectra and the analysis described, we determine an effective temperature of $T_\mathrm{eff} = $~\teff~K for $\tau$~Ceti.  

The effective temperature can also be calculated from the angular diameter and bolometric flux with the relation

\begin{equation}
    T_\mathrm{eff} = \left( \frac{4 F_\mathrm{bol}}{\sigma \theta_\mathrm{LD}^2} \right)^{1/4}
\end{equation}
where $F_\mathrm{bol}$ is the bolometric flux and $\sigma$ is the Stefan-Boltzmann constant.  For the bolometric flux, we used the value for $\tau$~Ceti determined by \citet{boyajian2013},  $F_\mathrm{bol} = (112.60000 \pm 0.0787) \times 10^{-8}$~erg~s$^{-1}$~cm$^{-2}$. This gives $T_\mathrm{eff} = 5370 \pm 20$~K.  The 1-$\sigma$ errors of this and the EXPRES $T_\mathrm{eff}$ overlap, showing agreement.

\subsection{Projected Rotational Velocity from Spectra} \label{sec:vsini_analysis}

During this spectral fitting, the ``total rotational broadening'' $v_{\mathrm{rot}}$ is the combined broadening from $v \sin{i}$ and macroturbulence, $v_\mathrm{mac}$.  The two different broadening kernels are similar, although $v \sin{i}$ can be thought of as being nearly constant on vertical slices parallel to the spin axis of the star, whereas $v_\mathrm{mac}$ is nearly constant in annuli centered on the star.  This is due to the varying radial and tangential components of the bulk motion caused by convection.  Microturbulence, Doppler broadening due to lower velocity thermal motions, is fixed at 1 \kms \ in this analysis.

\citet{brewer2016} derived a macrotubulence relation as a function of $T_{\mathrm{eff}}$ for both dwarf stars and subgiants from their sample of $\sim 1600$ stars observed with Keck HIRES.  They did this by assuming that the floor of the distribution of $v_{\mathrm{rot}}$ would be pole-on or non-rotating stars. The analysis then fixes the parameters derived from the first two stages, fixes $v_{\mathrm{mac}}$ using the relation, and fits for $v \sin{i}$.

$\tau$~Ceti was included as part of the \citet{brewer2016} analysis, but it was an outlier with all five spectra analyzed having total rotational broadening 1.5 $\sigma$ below the floor of the distribution.  Although the same procedure was used to analyze the EXPRES spectra, including the same line list, differences in the instrumental profile and spectral format can result in small differences between instruments.  In general, stellar parameters between stars in common between the two instruments agreed within the uncertainties.  The mean of the EXPRES measurements were $v_{\mathrm{rot}} = 2.14 \pm 0.05$ \kms.  This still falls below the mean of the macrotubulence relation of \citet{brewer2016}. The final fitting stage then resulted in $v \sin{i} = 0.08 \pm 0.03$ \kms.  Due to the uncertainty arising from the modeling, a more reasonable uncertainty would be 0.1 \kms, or about double the standard deviation in $v_{\mathrm{rot}}$.  We use $v \sin i$ = \vsini \ \kms \ for our further analyses of $\tau$~Ceti.

\begin{figure}
    \includegraphics[width=\linewidth]{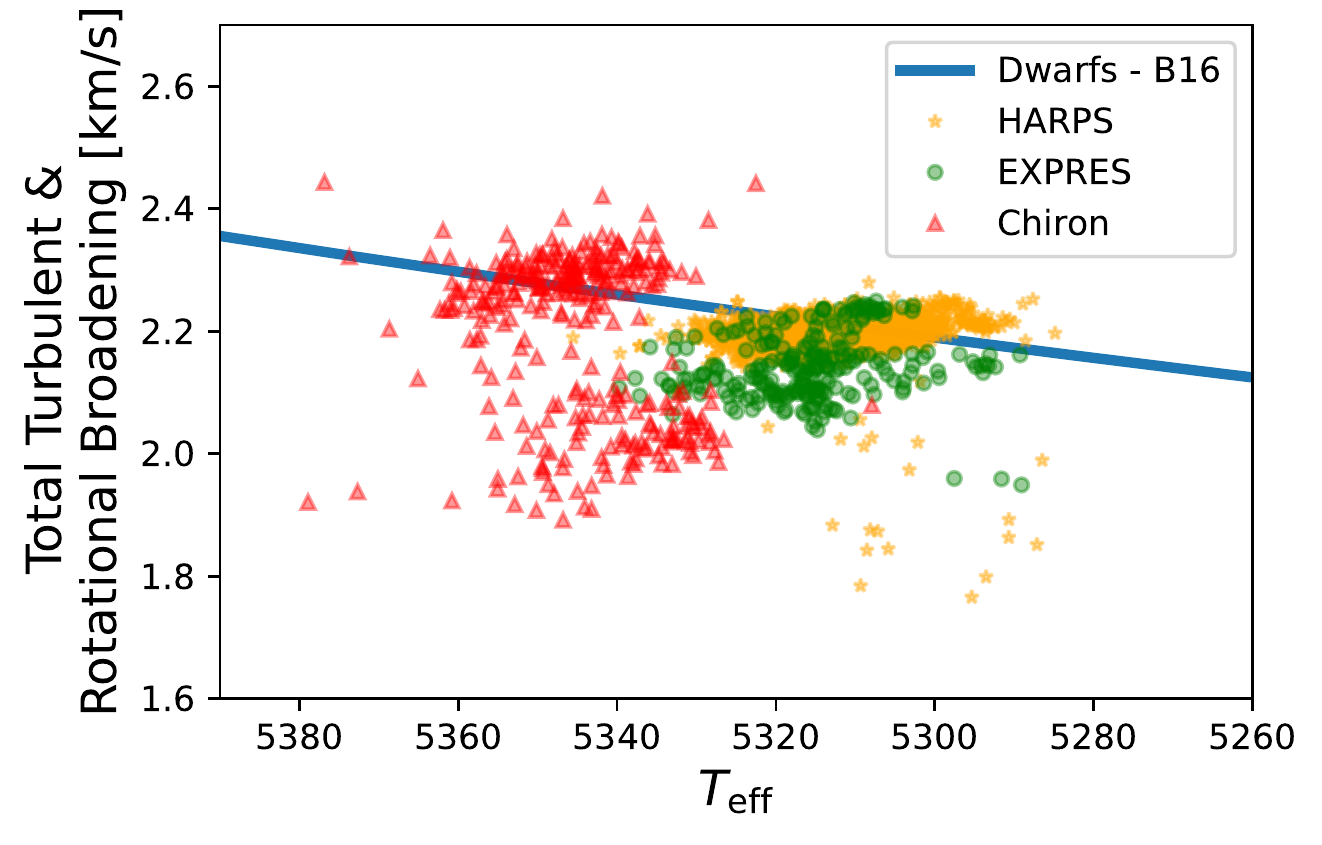}
    \caption{Total rotational broadening of 2934 spectra of $\tau$~Ceti from three different spectrographs.  The Chiron spectra (red) show a bimodal distribution caused by poor characterization of the instrumental profile when using the fiber slicer versus the slit, which has a higher resolution.  The dwarf macroturbulence relation from \citet{brewer2016} (blue line) shows that, for most of the spectra, there is no appreciable rotation beyond that likely due to macroturbulence, resulting in our low estimate of $v \sin i$ = \vsini \kms.} 
    \label{rotational_boradening}
\end{figure}

We performed an additional test to verify that the $v \sin{i}$ was consistent with zero.
We performed the same analysis described above on a total of 2,934 $\tau$~Ceti spectra from CHIRON 
\citep{tokovinin2013}, EXPRES, and HARPS.  No attempt was made to normalize the resulting parameters between the different spectrographs, since the parameters generally agreed to within the uncertainties.  The resulting rotational broadening was $v_{\mathrm{rot}} = 2.19 \pm 0.07 \ \kms$, falling below the relation from \citet{brewer2016} for $T_\mathrm{eff} \gtrsim 5280$ K, lower than the EXPRES value of $T_\mathrm{eff} =$ \teff \ K (see Figure \ref{rotational_boradening}).

\subsection{Age}

Age estimates for $\tau$~Ceti range from 4.4--12.4 Gyr \citep{lachaume1999,pijpers2003,difolco2004,mamajek2008,baum2022}.  The values from \citet{lachaume1999,pijpers2003, mamajek2008, baum2022} all depend upon estimates for the rotation period, $P_\mathrm{rot}$.  However, with a $v \sin i =$ \vsini \ \kms, which is consistent with little to no rotational velocity---potentially an indication of a pole-on orientation---we aim to investigate $\tau$~Ceti without the assumption that a periodic signal requiring rotation modulation has been detected.  \citet{difolco2004} does not use a rotation period but rather  a stellar evolution code that takes mass, luminosity, effective temperature, and initial chemical abundance as input.  They give an age estimate of 10 Gyr.

\subsection{Rotation Period and Inclination}
\label{analysis:inclination}

With gyrochronology, the age of the star,  rotation period, and  color index are related. As derived in \cite{barnes2007}, the age of the star can be expressed as follows:
\begin{equation}
    \label{eqn:age}
    \log{t} = \frac{1}{n} \left( \log{P_\mathrm{rot}} - \log{a} - b \log{((B-V) - 0.4)} \right),
\end{equation}
where $t$ is the age of the star in Myr, parameters $a$, $b$, and $n$ are constants, $P_\mathrm{rot}$ is the rotation period in days, and $B-V$ is the color index of the star.  The constants are determined by \citet{barnes2007} to be $a = 0.7725 \pm 0.011$, $b = 0.601 \pm 0.024$, and $n = 0.5189 \pm 0.0070$.
 $B-V = 0.72$ for  $\tau$~Ceti  \citep{ducati2002}.

Using the age estimate from \citet{difolco2004} in Equation \ref{eqn:age} and solving for the rotation period, we find $P_\mathrm{rot} = $ \prot~days.

To determine the inclination of $\tau$~Ceti, we use the range of rotation periods based on the age range from \citet{difolco2004}, the gyrochronology relationship given in Equation \ref{eqn:age}, the interferometrically determined stellar radius, and the spectroscopic $v \sin i$ to give an inclination of $7\pm7^\circ$.

With this range of inclinations, rotational variations may be visible, but only on the stellar limb.  We investigate possible indications of the rotation period of $\tau$~Ceti.  

We note that the periodograms in the next three subsections led to a few peaks nearly equal in power.  The strongest peaks for each are consistent with a nearly-pole-on orientation and we discuss those below.  

\ 

\ 

\subsubsection{MWO HK Project Rotation Period}

\begin{figure}
    \includegraphics[scale=0.575]{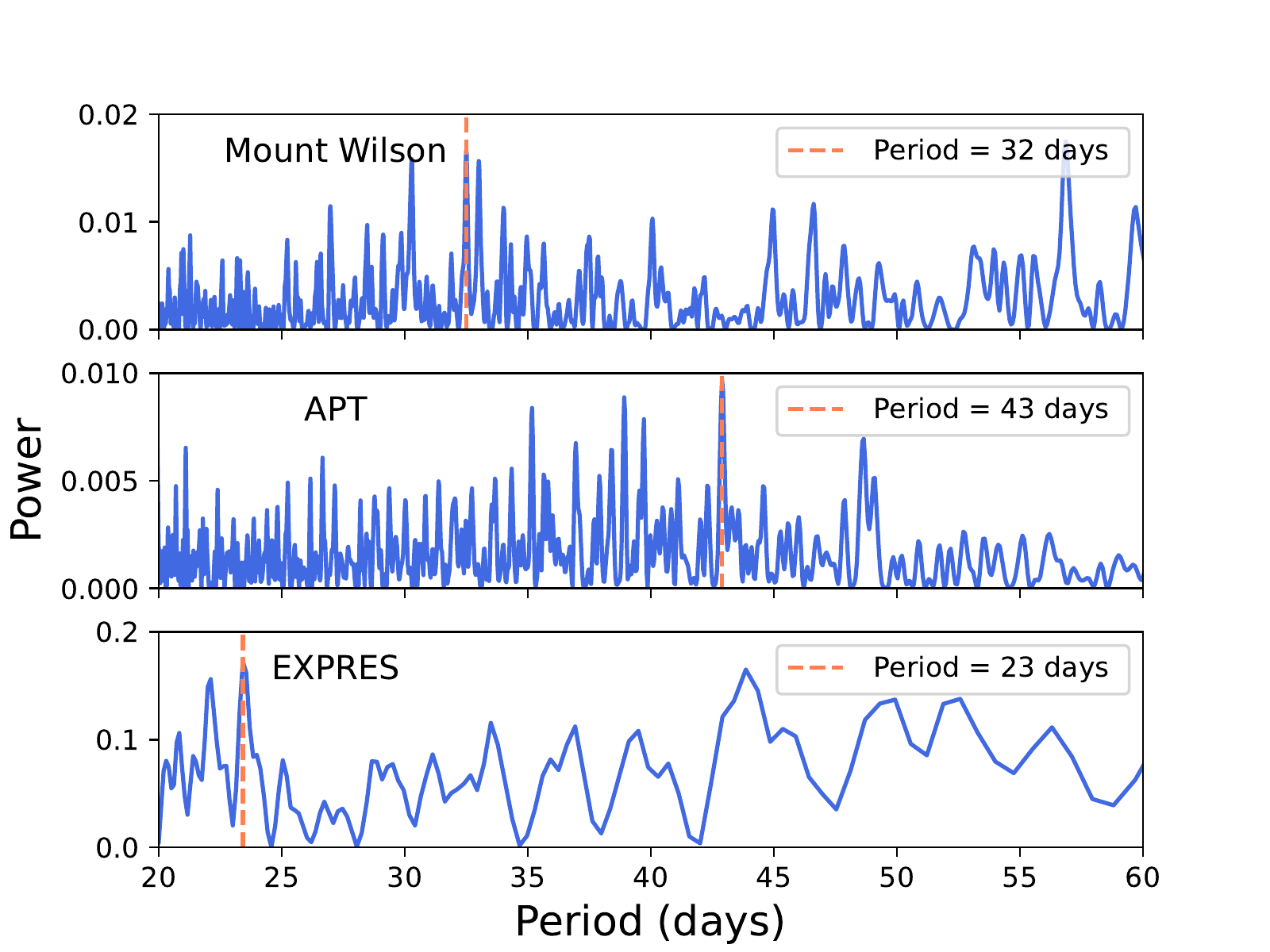}
    \caption{Periodogram created by chromospheric data from the Mount Wilson Observatory HK Project, photometric data from the Automatic Photoelectric Telescope at the Fairborn Observatory, and RV data from EXPRES. The red line shows the rotation period of the star at 32 days (MWO), 43 days (APT), and 23 days (EXPRES).  } 
    \label{mtwilson periodogram}
\end{figure}

We extracted a periodic signal from stellar chromospheric activity data from the MWO HK Project  using a Lomb-Scargle periodogram. From this periodic signal, which we assume is due to rotation, we determined the rotation period to be $32 \pm 9$ days, as seen in Figure \ref{mtwilson periodogram}. The errors were calculated using the bootstrap method, where we selected 1,784 points with replacement and found the best-fit $P_\mathrm{rot}$ 1,000 times.  The standard deviation of those 1,000 iterations is the 9-day error. This relatively large error of 9 days is consistent with the fact that other significant peaks seen in the periodogram are included within this range.  The false alarm probability of the peak in the periodogram is 0.004, indicating that the peak is statistically significant.  Since a period was detected with significance, an inclination slightly larger than $0^\circ$ is suggested, consistent with the results described above.  If this is the case, it implies that the periodic signal extracted from the data could be attributed to a rotational surface feature, such as starspots.  
This agrees with prior values from the literature of $34$ days \citep{baliunas1996}, which uses the previous processing of the MWO data set, and $34.5$~days \citep{saar1997}, which uses Ca II flux modulations. 

Combining our MWO rotation period and rotational velocity, we determined the star's inclination  to be \incMWO$^\circ$. Other peaks within the errors of the rotation period lead to inclinations within the errors of \incMWO$^\circ$. The relatively weak signals in the periodogram are consistent with the pole-on inclination of the star, as the low inclination makes the rotation difficult to detect. Other comparatively strong peaks, such as that at $45$~days, are not far from the $P_\mathrm{rot}$ we derived with an age of $10$~Gyr.

\ 

\subsubsection{APT Rotation Period}

We performed a periodogram analysis of the ground-based APT data to determine the rotation period of $\tau$~Ceti,  $P_\mathrm{rot} = 43 \pm 7$ days.  The errors were determined through the bootstrap method where 1,000 light curves containing 1,369 points (the number of points in the observed APT light curve)   were randomly chosen with replacement. The power in the periodogram is very low, and there are several comparable peaks at other periods, including 35 days, which is consistent with our MWO $P_\mathrm{rot}$. Using our radius and $v \sin i$ values, the inclination is calculated to be $6 \pm 6^{\circ}$ for a rotation period of $43 \pm 7$ days.   \par

While these results are consistent with our analyses described above, we note that the peaks in the periodogram are weak.  The false-alarm probability of all peaks in this periodogram is near 1, which is consistent with a pole-on star and little rotational variation.     \par

\subsubsection{EXPRES RV Rotation Period}

We performed a periodogram analysis on the EXPRES RVs to determine a rotation period.  There were, however, no strong signals in this data set (see Figure \ref{mtwilson periodogram}).  The strongest peak in the EXPRES data is found at 23 days and gives $i = 3 \pm 3^\circ$.  It is possible with a longer temporal baseline of monitoring that a signal associated with the rotation period may be detected with more significance.

\subsection{Surface Gravity}

Like the values for $v \sin i$ and $T_\mathrm{eff}$, surface gravity, $\log g$, is determined from EXPRES data and model fitting described above in Section \ref{sec:vsini_analysis}.  These give a value of $\log g = $ \logg.  

\subsection{Mass}
\label{discussion:mass}

 From $\log g = $ \logg \ and the interferometric radius, we find a mass of \masslogg \ $M_\odot$.

A star's mass can also be calculated using  asteroseismic scaling relations for $\nu_\mathrm{max}$ and $\Delta\nu$:
\begin{equation}
\label{eqn:vmax}
    \nu_\mathrm{max} = \left(\frac{M}{M_{\odot}}\right)\left(\frac{R}{R_{\odot}}\right)^{-2}\left(\frac{T}{T_{\odot}}\right)^{-1/2} \nu_{\mathrm{max}\odot}
\end{equation}
and \begin{equation}
\label{eqn:deltanu}
    \Delta\nu = \left(\frac{M}{M_{\odot}}\right)^{1/2}\left(\frac{R}{R_{\odot}}\right)^{-3/2} \Delta\nu_{\odot},
\end{equation}
where $M$, $R$, and $T$ are the mass, radius, and temperature of $\tau$~Ceti, respectively. $M_\odot$, $R_\odot$, and $T_\odot$ are the solar values for these parameters.
We used solar asteroseismic values  from \cite{huber2011} and $\tau$~Ceti's asteroseismic values from
\citet{teixeira2009}: $\nu_\mathrm{max} = 4100 \ \mu$Hz and $\Delta\nu = 169 \ \mu$Hz. 
Thus, from the frequency of maximum power and effective temperature \teff K and the limb-darkened radius of \radiusLD~mas determined above, we calculate $\tau$~Ceti's mass to be \massvmax~$M_\odot$. From the large frequency separation and the limb-darkened radius, the mass is calculated to be \massdv~$M_\odot$, which is within the $1\sigma$ error of our mass derived above. These values are also consistent with literature values, which average around $0.78 \ M_\odot$ (see values in Table \ref{literature} in the Appendix). The mass errors were calculated using the standard deviation of masses calculated by randomly picking a value from the Gaussian distribution of the other terms' errors. \par

\section{Dynamical Stability}
\label{dynamics}

The new constraints on the inclination of the stellar spin axis presented in this work have significant consequences for RV planets detected in the $\tau$~Ceti system. \citet{feng2017} and \citet{tuomi2013} reported the discovery of four exoplanets orbiting $\tau$~Ceti, with orbital periods in the range of 20--636~days and semi-major axes of 0.133--1.334~AU, interior to the debris disk reported by \citet{macgregor2016}.  The planets are reported to have masses ($m \sin i$) in the range of 1.75--3.93~$M_\oplus$.  The reported planetary masses are minimum masses for the specific case of co-planar orbits that are aligned with the line of sight ($i = 90^\circ$). It has been suggested that the planets are rocky and that additional planets may exist in the system within the orbital gaps \citep{dietrich2021}. However, assuming that the planetary orbits are co-planar \citep{masuda2020} and possess a low obliquity with respect to the stellar spin axis \citep{albrecht2022}, which is likely given the age of the system, the results presented in this paper imply a dramatic increase in the planetary masses. For example, inclinations of $i = 6^\circ$ and $i = 1^\circ$ increase the planetary masses by factors of $\sim$10 and $\sim$60, respectively. This means that the four known planets are likely substantially more massive than the minimum masses provided by \citet{feng2017}, such that they are not terrestrial in nature with masses that exceed that of Uranus and Neptune.

Given the planetary mass increase, we conducted a suite of dynamical simulations to test the dynamical integrity of the system. The N-body integrations were performed via the Mercury Integrator Package \citep{chambers1999} using methodology similar to that described by \citet{kane2015,kane2016,kane2019}. Based upon our inclination range, we investigated orbital inclinations in the range 1--10$\degr$ in steps of 1$^\circ$, adjusting the \citet{feng2017} planetary masses accordingly. Each simulation was run for $10$~Myr. Based on the inner planet orbital period of 20~days, we adopted a conservative time step for the simulations of 0.1 days to assure perturbative reliability. As quantified by \citet{duncan1998}, the time step should be, at minimum, 1/20 of the shortest orbital period; our time step is 1/200. Our simulations show that there is a significant transfer of angular momentum that occurs between the planets with all simulations that increases the eccentricity range of the planets compared to the initial values, suggesting that long-term stability is unlikely to be viable within the tested inclination regime. Importantly, the system is rendered unstable in less than $0.1$~Myr for the case of $i = 1^\circ$ of both the star and the planets, implying that the planetary architecture described by \citet{feng2017} cannot exist for that inclination scenario.   Due to uncertainties in the orbital parameters, there is limited reliability in the dynamical simulation results when integrating beyond $10$~Myr.  For simulations run for $10$~Myr with an inclination of $7^\circ$, the planets are nearing the instability threshold suggesting that, given more time, the system would also become unstable.

A face-on inclination for the $\tau$~Ceti system increases its viability as a direct imaging target from the perspective of planetary orbit visibility \citep{kane2013,dulz2020}. Direct imaging observations of the system thus far have placed upper limits on the presence of giant planets at large separations from the host star \citep{pathak2021}. Further observations with the Roman Space Telescope will provide valuable additional constraints on possible giant planets present in the system \citep{turnbull2021}.

\section{Conclusion}

We revised stellar parameters for $\tau$~Ceti with the assistance of new optical interferometric and spectroscopic data.  Building upon fundamental observations, we formed a consistent picture of $\tau$~Ceti that shows it is nearly pole-on.  As a result of the inclination, there are difficulties in reliably determining a rotation period and detecting planets with any method other than potentially future direct imaging.  The orientation of the stellar rotation axis makes the detection of surface features like starspots difficult because their rotational modulations will only be detectable should they be nearly equatorial to allow for rotation over the stellar limb.  This alignment also makes observing transits or RV shifts unlikely, unless the planets are significantly misaligned with the stellar rotation axis.    

Because the potential planets described by \citet{feng2017} fall between 0.133-1.33 AU, we assumed that their orbital plane would be aligned with the stellar rotation axis in our analysis in Section \ref{dynamics}.  
 While still within 3-$\sigma$ errors, our nearly pole-on inclination of $7 \pm 7^\circ$ differs from the debris disk inclination of $35 \pm 10^\circ$ \citep{lawler2014}, which used observations from the Herschel satellite and had a beam size comparable to the size of the debris disk.  A more recent study with ALMA data \citep{macgregor2016} assumed the inclination of $35^\circ$ from \citet{lawler2014} and did not provide an independent fit to either the ALMA or Herschel data. If the difference in inclinations is real, this could imply that the disk and potential planets are misaligned with the star, or---since the debris disk result agreed with previous stellar inclination measurements of $0-40^\circ$ \citep{greaves2004}---it could suggest that a more accurate measurement of the debris disk inclination would be consistent with our pole-on stellar inclination.     
The possible misalignment between the stellar rotation axis and the debris disk potentially indicates a complicated formation scenario.

More interferometric observations would allow for imaging of the stellar surface  potentially to see the rotation of surface structures, which may not modulate photometric or spectroscopic observations.  Our current data, however, are not sufficient for imaging, as it is too limited in $uv$ coverage and time.    A new set of data obtained during a single stellar rotation would allow for the unambiguous confirmation  of the stellar inclination and help place limits on the spottedness of the stellar surface. Observations taken throughout the stellar rotation, maximizing the $uv$ coverage across the stellar surface and with sufficient resolution to resolve surface features can be obtained with the six-telescope beam combiners at the CHARA Array.  While MIRC-X can provide these capabilities in $H$-band, the Stellar Parameters and Images with a Cophased Array (SPICA) beam combiner \citep{mourard2022} will operate in optical wavelengths and will soon be available to the public.  SPICA will provide the opportunity to achieve higher-resolution images of stellar surfaces than is currently possible. Such a precise new data set is needed to improve upon our results and is necessary for characterizing both the star and any planets it hosts.

\section*{ACKNOWLEDGEMENTS}

These results made use of the Lowell Discovery Telescope at Lowell Observatory. Lowell is a private, non-profit institution dedicated to astrophysical research and public appreciation of astronomy and operates the LDT in partnership with Boston University, the University of Maryland, the University of Toledo, Northern Arizona University and Yale University.  Lowell Observatory sits at the base of mountains sacred to tribes throughout the region. We honor their past, present, and future generations, who have lived here for millennia and will forever call this place home.
Support for the design and construction of EXPRES was supported by the National Science Foundation (NSF) MRI-1429365, NSF ATI-1509436 and Yale University. We gratefully acknowledge support to carry out this research from NSF 2009528, NSF 1616086, NASA 17-XRP17 2-0064, the Heising-Simons Foundation, and an anonymous donor in the Yale alumni community.  
A portion of the CHARA Array time was granted through the NOIRLab community-access program (NOIRLab Prop. ID: 2021B-0153; PI: R.\ Roettenbacher). The CHARA Array is supported by the National Science Foundation under Grant No. AST-1636624 and AST-2034336, the GSU College of Arts and Sciences, and the GSU Office of the Vice President for Research and Economic Development. CHARA telescope time was granted by NOIRLab through the Mid-Scale Innovations Program (MSIP). MSIP is funded by NSF.
MIRC-X received funding from the European Research Council (ERC) under the European Union’s Horizon 2020 research and innovation programme (Grant No. 639889). This research has made use of the Jean-Marie Mariotti Center \texttt{Aspro} service\footnote{Available at http://www.jmmc.fr/aspro}.
The APT photometric data were supported by NASA, NSF, Tennessee State University, and the State of Tennessee through its Centers of Excellence program.  
This research has made use of the SIMBAD database, operated at CDS, Strasbourg, France.
The HK\_Project\_v1995\_NSO data derive from the Mount Wilson Observatory HK Project, supported by both public and private funds through the Carnegie Observatories, the Mount Wilson Institute, and the Harvard-Smithsonian Center for Astrophysics starting in 1966 and continuing for over 36 years.
RMR acknowledges support from the Yale Center for Astronomy \& Astrophysics (YCAA), the Heising-Simons Foundation, and NASA EPRV 80NSSC21K1034. JDM acknowledges funding for the development of MIRC-X (NASA-XRP NNX16AD43G, NSF-AST 1909165). SK acknowledges support from ERC Consolidator Grant GAIA-BIFROST (Grant Agreement ID 101003096) and STFC Consolidated Grant (ST/V000721/1).  JL acknowledges support from NSF award AST-2009501. JMB acknowledges support from NASA grant 80NSSC21K0009 and NASA-XRP 80NSSC21K0571.

\facilities{CHARA, DCT, TSU:APT}

\software{Astropy \citep{astropy:2013, astropy:2018}, Astroquery \citep{astroquery}, NumPy \citep{numpy}, SciPy \citep{scipy}, Matplotlib \citep{matplotlib}. }

\clearpage
\appendix

\section{Literature Table}
\label{lit table}

For a detailed comparison of the values determined by the methods described above, we include the stellar parameters determined by previous studies.  In Table \ref{literature}, we include literature values and notes on how those values were obtained.

\begin{turnpage}
\begin{deluxetable}{l c c c c c c c c}
\label{literature}
\tablecaption{Recent $\tau$~Ceti Literature Values}
\tablehead{
 \colhead{Reference} & \colhead{Temperature} & \colhead{Mass} & \colhead{Radius} & \colhead{Luminosity } & \colhead{Age } & \colhead{Angular } & \colhead{Angular } & \colhead{Method}\\
 & \colhead{(K)} & \colhead{($M_\odot$)} & \colhead{($R_\odot$)} & \colhead{($L_\odot$)} & \colhead{(Gyr)} & \colhead{Diameter} & \colhead{Diameter} & \colhead{}\\
 & & & & & & \colhead{(mas; UD)} & \colhead{(mas; LD)} & 
 }

\startdata
\hline
This work & \teff & \masslogg & \radiusLD & \lumin & -- & \thetaUD & \thetaLD & interferometry + spectroscopy\\
\citet{baum2022} & 5333 & 0.990 & -- & -- & 12.4 & -- & -- & spectroscopy \\
 \cite{tabernero2021} & $5400 \pm 60$ & $0.760 \pm .017$ & $0.750 \pm .015$ & -- & -- &  -- & -- & spectroscopic modelling\\
 \cite{esposito2020}& 5750 & $0.85 \pm .01$ & 0.75 & $0.56 \pm 0.23$ & -- &  -- & -- &  optical photometry\\
 \cite{rains2020}& $5347 \pm 18$ & -- & $0.796 \pm .004$ & $0.47 \pm 0.01$ &  -- & $2.005 \pm 0.011$ &  $2.054 \pm 0.011$  &  interferometry + flux\\
 \cite{chaplin2019}& 5290 & 0.79 & 0.85 & 0.51 &  -- &  -- &  -- & spectroscopy\\
 \cite{kervella2019}& -- & $0.900 \pm .045$ & $0.751 \pm .014$ & -- &  -- &  -- & -- &  isochrone fitting   + surface \\
&  &  &  &  &   &   &  & brightness-color relation\\
 \cite{france2018}& $5340 \pm 36$ & -- & $0.793 \pm .036$ & -- & -- &  -- &  -- &  spectral type\\
 \cite{fuhrmann2017}& -- & 0.78 & -- & -- &  -- &  -- & -- &  stellar evolutionary track\\
 \cite{brewer2016}& $5344 \pm 60$ & $0.78 \pm .02$ & 0.82 & -- &  -- &  -- & -- &  spectroscopy\\
 \cite{heiter2015}& $5326 \pm 45$ & $0.71 \pm .03$ & -- & $0.447 \pm 0.005$ &  -- &  -- & -- &  spectroscopy + isochrone\\
 \cite{pagano2015}& 5387 & 0.78 & 0.69 & 0.504 &  -- &  -- &  -- & spectroscopy\\
  \cite{baines2014} & -- & --& --& --& --& $1.952 \pm 0.003$  & $2.072 \pm 0.010$ & interferometry \\
 \cite{absil2013} & -- & --& --& --& --& -- & $2.015 \pm 0.004$ & interferometry \\
 \cite{boyajian2013}& $5290 \pm 39$ & 0.733 & $0.815 \pm .012$ & $0.4674 \pm 0.0007$ &  -- &  -- & -- &parallax/flux + isochrone\\
 \cite{jofre2013}& $5414 \pm 21$ & $0.78 \pm .01$ & 0.69 & -- &  -- &  -- & -- &spectroscopy\\
 \cite{tang2011}& 5409 & 0.775 & 0.790 & 0.47985 &  -- &  -- & -- & asteroseismology model 1\\
 \cite{tang2011}& 5387 & 0.785 & 0.793 & 0.47612 &  -- &  -- & -- & asteroseismology model 2\\
 \cite{tang2011}& $5264 \pm 100$ & -- & 0.87 & $0.52 \pm 0.03$ &  -- &  -- & -- & spectroscopy\\
 \cite{tang2011}& $5525 \pm 12$ & -- & 0.77 & $0.50 \pm 0.006$ &  -- &  -- & -- & spectroscopy + interferometry\\
 \cite{bruntt2010}& $5383 \pm 47$ & $0.79 \pm .03$ & $0.794 \pm .005$ & $0.47 \pm 0.02$ &  -- &  -- &  -- & interferometry + photometry\\
 \cite{teixeira2009}& 5418 & $0.783 \pm .012$ & $0.793 \pm .004$ & $0.488 \pm 0.010$ &  -- &  -- &  -- & parallax + asteroseismology\\
\citet{mamajek2008} & -- & -- & -- & -- & 5.8 & -- & -- & activity-rotation \\
 \cite{sousa2008}& $5310 \pm 17$ & 0.627 & 0.62 & $0.495 \pm 0.003$ &  -- &  -- &  -- & spectroscopy\\
 \cite{difolco2007}& 5400 & 0.72 & $0.790 \pm .005$ & -- &  -- &  -- &  -- & parallax\\
 \cite{difolco2004}& $5264 \pm 100$ & $0.85 \pm .14$ & $0.806 \pm .013$ & -- &  -- &  $2.005 \pm 0.034$ &  -- & spectrophotometry + interferometry\\
 \cite{difolco2004}& 5377 & 0.83 & 0.821 & -- &   10 &  -- &  -- &  stellar evolutionary track\\
 \cite{pijpers2003}& $5264 \pm 100$ & 0.50 & $0.773 \pm .004$ & $0.52 \pm 0.03$ &  9-10 & $1.933 \pm 0.009$  &  $1.971 \pm 0.009$ & interferometry + spectroscopy\\
\enddata
 \tablecomments{Some references are listed multiple times, as multiple methods were used to determine the stellar parameters}
\end{deluxetable}
\end{turnpage}

\clearpage
\bibliography{tauCet.bib}

\end{document}